\documentclass[aps,superscriptaddress]{revtex4}
\usepackage{amssymb}
\usepackage{amsmath,bm}
\usepackage[normalem]{ulem}
\usepackage[dvips]{color}
\usepackage{subfigure}
\usepackage[english]{babel}
\usepackage{multirow}
\usepackage[mathcal]{eucal}
\usepackage{rotating}
\usepackage{graphicx}
\usepackage{dcolumn}
\usepackage{bm}

\newcommand{\vd}{v_{\chi}}
\newcommand{\vdbf}{\boldsymbol{\mathit{v}}_{\chi}}

\newcommand{\vebf}{\boldsymbol{\mathit{v}}_{\rm E}}

\newcommand{\vmin}{v_{\rm min}}

\renewcommand\sout{\bgroup \color{red} \ULdepth=-.5ex \ULset}

\begin{document}

\title{Form Factor Effects in the Direct Detection of Isospin-Violating Dark Matter}
\author{Hao Zheng$^1$, Zhen Zhang$^1$ and Lie-Wen Chen\footnote{%
Corresponding author (email: lwchen$@$sjtu.edu.cn)}}
\affiliation{Department of Physics and Astronomy and Shanghai Key Laboratory for
Particle Physics and Cosmology, Shanghai Jiao Tong University, Shanghai 200240, China}
\affiliation{Center of Theoretical Nuclear Physics, National Laboratory of Heavy Ion
Accelerator, Lanzhou 730000, China}
\date{\today}

\begin{abstract}
Isospin-violating dark matter (IVDM) provides a possible mechanism to
ameliorate the tension among recent direct detection experiments.
For IVDM, we demonstrate that the results of direct detection experiments
based on neutron-rich target nuclei may depend strongly on the density dependence
of the symmetry energy which is presently largely unknown and controls
the neutron skin thickness that reflects the relative difference of neutron
and proton form factors in the neutron-rich nuclei. In particular, using
the neutron and proton form factors obtained from Skyrme-Hartree-Fock
calculations by varying the symmetry energy within the uncertainty region
set by the latest model-independent measurement of the neutron skin thickness
of $^{208}$Pb from PREX experiment at JLab, we find that, for IVDM with
neutron-to-proton coupling ratio fixed to $f_n/f_p=-0.7$, the form factor
effect may enhance the sensitivity of Xe-based detectors (e.g., XENON100
and LUX) to the DM-proton cross section by a factor of $3$ in the DM mass
region constrained by CMDS-II(Si) and even by more than an order of magnitude
for heavy DM with mass larger than $80$ GeV, compared with the results using
the empirical Helm form factor. Our results further indicate
that the form factor effect can significantly modify the recoil spectrum of
Xe-based detectors for heavy IVDM with $f_n/f_p=-0.7$.
\end{abstract}

\maketitle

\section{Introduction}
\label{sec:intro}

The possible existence of dark matter (DM) is one of the most intriguing
aspects of modern particle physics, astrophysics and cosmology. In order
to survey the nature of DM, a number of observations and experiments have
been conducted or are underway around the world. The most recent cosmological
results based on {\it Planck} measurements of the cosmic microwave background
(CMB) temperature and lensing-potential power spectra indicate that DM
comprises about $27\%$ of the energy density of the Universe which also
includes about $5\%$ baryon matter and about $68\%$ dark energy~\cite{Plk13}.
Many theories beyond the Standard Model of particle physics predict natural candidates for
DM, e.g., the weakly interacting massive particles (WIMPs) which are a class
of hypothetical stable neutral particles with a huge range in masses from $1$
GeV to $100$ TeV and interaction cross sections with normal matter (proton)
from $10^{-40}$ to $10^{-50}$ cm$^{2}$~\cite{Ste85,Jung96}. In terrestrial laboratory,
DM might be directly detected through their elastic scattering off nuclei
in particle detectors~\cite{Goodman85}. A number of underground DM direct
detection experiments have been performed, and among them an excess of events
over the expected background has been observed by CoGeNT~\cite{CoGeNT11},
DAMA~\cite{DAMA09}, CRESSTII~\cite{CRESST12} as well as in the recent results
presented by the CDMS-II(Si) collaboration~\cite{CDMS13v1,CDMS13v2}. However, these
results are in strong tension with the constraints set by some other experimental
groups like XENON100~\cite{XENON100v11,XENON100v12}, LUX~\cite{LUX13} and
SuperCDMS(Ge)~\cite{SCDMS14}, leaving a confusing situation for the community. This
has led to a number of attempts trying to explain the discrepancy by considering
atomic uncertainties~\cite{Sav11} or different mechanisms that deviate
from standard assumptions about DM interactions or its astrophysical
distributions~\cite{Tuc01,Fra12,Mao13,Cot13}.

Isospin-Violating Dark Matter (IVDM) provides a very promising mechanism to
reconcile the tension among different experiments~\cite{Kur04,Giu05,Cha10,Feng11,Feng13a,Feng13b,Nag13,Vin13}.
Within the IVDM framework, DM is assumed to couple differently with protons
and neutrons, and this assumption of the isospin violation has been supported
by a number of theoretical works~\cite{Fra11,Cli11,Del12,He12,Gao13,Oka13} based on the
particle physics point of view. Many parameters need to be specified in the standard
method of analyzing DM direct detection experiments~\cite{Lew96} and recently
Frandsen~{\it et al.}~\cite{Fran13} presented a systematic discussion on the possible
ways to ameliorate the tension among different experiments. They found that the
tension between the CDMS-II(Si) results and the XENON100 bounds is independent of
the astrophysical uncertainties concerning the DM halo and any momentum- and
velocity-dependence of the cross section in particle physics, but it can be
largely ameliorated or even resolved within the framework of IVDM.

Besides the uncertainties in astrophysics and particle physics concerning
the interaction of a DM particle scattering off a single nucleon mentioned
above, the uncertainties in nuclear physics describing how the struck nucleon is
distributed inside the nucleus may also play an important role in interpreting DM
signals. This is because in DM direct detection experiments, the nuclear form
factors are generally applied to describe the DM-nucleus cross section and the
bounds on DM-nucleon cross section are then obtained accordingly. In particular,
the empirical Helm form factor extracted from nuclear charge distributions~\cite{Helm56,Duda07}
has been commonly adopted in current direct detection experiments. However, the
DM-nucleus interaction should be in principle described by using the form factors
of both proton and neutron distributions in the nuclei rather than the charge
distributions since the DM particles actually interact with the protons and
neutrons in the nuclei. Within the framework of relativistic mean field (RMF)
model, Chen~{\it et al.}~\cite{ChenYZ11} derived the nuclear form factor for the
spin-independent scattering between the WIMPs and nucleus, and they found that
the results can deviate from the empirical Helm form factor by $15\%$ to $25\%$
in a large range of recoil spectrum of $0 \sim 100$ keV. A recent work by
Co' {\it et al.}~\cite{Co12} suggests that the use of different distributions
for protons and neutrons instead of the commonly used empirical charge
distributions for a target nucleus could be important in interpreting DM
signals, especially if IVDM is considered.

For stable nuclei, the proton distribution can be well determined from the
charge distribution which can be accurately measured with electron
scattering~\cite{Hof56,Frois77,Dej87}. In contrast, the neutron distribution is usually
determined from hadron scattering experiments and the results are generally
highly model dependent due to the unclear non-perturbative strong interaction~\cite{PREX12}.
Recently, the Lead Radius Experiment (PREX) collaboration at Jefferson
Laboratory (JLab) published their results on the measurement of the parity-violating
cross section asymmetry in the elastic scattering of polarized electrons from
$^{208}{\rm Pb}$~\cite{PREX12}, which provides a model-independent determination of the
neutron density distributions in $^{208}{\rm Pb}$. The PREX measurement leads to a value of
$\sqrt{\left\langle r_{\rm n}^2 \right \rangle} =5.78_{-0.18}^{+0.16} \,\, {\rm fm}$
for the rms radius of the neutron distributions for $^{208}{\rm Pb}$, implying a
large neutron skin thickness, i.e.,
$\Delta r_{\rm{np}}=\sqrt{\left\langle r_{\rm n}^2 \right \rangle} - \sqrt{\left\langle r_{\rm p}^2 \right \rangle} = 0.33_{-0.18}^{+0.16} \,\, {\rm fm}$ by assuming a point-proton rms radius of $5.45$ fm~\cite{Ong10}. One can
see that the obtained value of $\Delta r_{\rm{np}}$ has a large error, indicating
a large uncertainty of the neutron distribution relative to the proton distribution
in $^{208}{\rm Pb}$.

Theoretically, it has been established~\cite{Horo01,Brown01} that the $\Delta r_{\rm{np}}$ is
intimately related to the nuclear matter symmetry energy which characterizes the isospin
dependent part of the equation of state (EOS) of asymmetric nuclear matter (ANM)~\cite{LCK08}.
In particular, it has been shown recently that the $\Delta r_{\rm{np}}$ of heavy
nuclei is uniquely determined by the density slope $L(\rho_{\rm c})$ of the symmetry
energy at a subsaturation cross density $\rho_{\rm c}\approx0.11$ fm$^{-3}$~\cite{Zha13}.
These features imply that the uncertainties of $\Delta r_{\rm{np}}$, especially
the neutron distributions, predicted by various nuclear models are essentially
due to our poor knowledge about the symmetry energy. The symmetry energy is of
critical importance for understanding not only the structure and reaction of
radioactive nuclei, but also a number of interesting issues in astrophysics,
such as the structure of neutron stars and the mechanism of supernova explosions,
and has become a hot topic in current research frontiers of nuclear physics
and astrophysics~\cite{EPJAEsym14}. The determination of the symmetry energy
provides a strong motivation for studying isospin-dependent phenomena with
radioactive nuclei at a number of new/planning rare isotope beam facilities around
the world, such as CSR/Lanzhou and BRIF-II/Beijing in China, RIBF/RIKEN in Japan,
SPIRAL2/GANIL in France, FAIR/GSI in Germany, SPES/LNL in Italy, RAON in Korea,
and FRIB/NSCL and T-REX/TAMU in USA.
In this work, using the proton and neutron form factors obtained from
Skyrme-Hartree-Fock calculations by varying the symmetry energy slope parameter
$L(\rho_c)$ within the uncertainty region set by the PREX experiment,
we investigate the form factor effects in the direct detection of dark
matter.

This article is organized as follows. We briefly introduce in Sec.~\ref{sec:model}
the theoretical models and methods used in the present work, and then present
the results and discussions in Sec.~\ref{sec:sym}. Finally, a conclusion is
given in Sec.~\ref{sec:conclusion}.

\section{Models and Methods}
\label{sec:model}

\subsection{The symmetry energy and Skyrme-Hartree-Fock approach}
The EOS of isospin asymmetric nuclear matter, defined by its binding energy per nucleon,
can be well approximated by
\begin{equation}
E(\rho ,\delta )=E_{0}(\rho )+E_{\mathrm{sym}}(\rho )\delta ^{2}+\mathcal{O}(\delta
^{4}),  \label{EOSANM}
\end{equation}%
in terms of baryon density $\rho=\rho_{\rm p}+\rho_{\rm n}$ and isospin asymmetry
$\delta=(\rho_{\rm n}-\rho_{\rm p})/\rho$ with $\rho_{\rm p}$ and $\rho_{\rm n}$ denoting
the proton and neutron densities, respectively. $E_{0}(\rho ) = E(\rho ,\delta=0)$ corresponds to
the EOS of symmetric nuclear matter, and the nuclear symmetry energy can be expressed as
\begin{equation}
E_{\mathrm{sym}}(\rho )=\left.\frac{1}{2!}\frac{\partial ^{2}E(\rho ,\delta )}{%
\partial \delta ^{2}}\right|_{\,\delta =0}.  \label{Esym}
\end{equation}
There are no odd-order $\delta$ terms in Eq.~(\ref{EOSANM}) due to the
exchange symmetry between protons and neutrons (isospin symmetry) in nuclear matter.
Neglecting the contribution from higher-order terms in Eq.~(\ref{EOSANM}) leads to
the well-known empirical parabolic law for EOS of ANM, which has been verified by
all many-body theories to date, at least for densities up to moderate values~\cite{LCK08}.

Furthermore, around a reference density $\rho_{\rm r}$, the symmetry
energy $E_{\mathrm{sym}}(\rho)$ can be expanded as
\begin{equation}
E_{\mathrm{sym}}(\rho )=E_{\text{\textrm{sym}}}({\rho _{\rm r}})+\frac{L({\rho _{\rm r}})}{3} \left(%
\frac{\rho -{\rho _{\rm r}}}{{\rho _{\rm r}}}\right)+\mathcal{O}\left(\frac{\rho -{\rho _{\rm r}}}{{\rho _{\rm r}%
}}\right)^{2},
\end{equation}
with
\begin{equation}
L({\rho _{\rm r}})=\left.3\rho _{\rm r}\frac{\partial E_{\mathrm{sym}}(\rho )}{\partial \rho }\right|_{\,\rho
=\rho _{\rm r}}.  \label{L}
\end{equation}
The slope parameter $L(\rho_{\rm r})$ characterizes the density dependence of
symmetry energy around $\rho_{\rm r}$. It has been shown~\cite{Zha13} that the
neutron skin thickness $\Delta{r_{\rm np}}$ of heavy nuclei is uniquely fixed
by the density slope $L(\rho_c)$ at a subsaturation cross density
$\rho_c \approx 0.11 $fm$^{-3}$ which roughly corresponds to the average
density of the nuclei.

For the calculations of finite nuclei, we use in the present work the
standard Skyrme-Hartree-Fock (SHF) approach in which the nuclear effective
interaction is taken to have a zero-range, density- and momentum-dependent
form, i.e.~\cite{Cha97},
\begin{eqnarray}
V_{12}(\mathbf{R},\mathbf{r}) &=&t_{0}(1+x_{0}P_{\sigma })\delta (\mathbf{r})
\notag \\
&+&\frac{1}{6}t_{3}(1+x_{3}P_{\sigma })\rho ^{\sigma }(\mathbf{R})\delta (%
\mathbf{r})  \notag \\
&+&\frac{1}{2}t_{1}(1+x_{1}P_{\sigma })(K^{^{\prime }2}\delta (\mathbf{r}%
)+\delta (\mathbf{r})K^{2})  \notag \\
&+&t_{2}(1+x_{2}P_{\sigma })\mathbf{K}^{^{\prime }}\cdot \delta (\mathbf{r})%
\mathbf{K}  \notag \\
&\mathbf{+}&iW_{0}(\mathbf{\sigma }_{1}+\mathbf{\sigma }_{2})\cdot \lbrack
\mathbf{K}^{^{\prime }}\times \delta (\mathbf{r})\mathbf{K]},  \label{V12Sky}
\end{eqnarray}%
with $\mathbf{r}=\mathbf{r}_{1}-\mathbf{r}_{2}$ and $\mathbf{R}=(\mathbf{r}%
_{1}+\mathbf{r}_{2})/2$. In the above expression, the relative momentum
operators $\mathbf{K}=(\mathbf{\nabla }_{1}-\mathbf{\nabla }_{2})/2i$ and $%
\mathbf{K}^{\prime }=-(\mathbf{\nabla }_{1}-\mathbf{\nabla }_{2})/2i$ act on
the wave function on the right and left, respectively. The quantities $%
P_{\sigma }$ and $\sigma _{i}$ denote, respectively, the spin exchange
operator and Pauli spin matrices.

The Skyrme interaction in Eq.~(\ref{V12Sky}) includes totally $10$ parameters,
i.e., the $9$ Skyrme interaction parameters $\sigma $, $t_{0}-t_{3}$, $x_{0}-x_{3}$,
and the spin-orbit coupling constant $W_{0}$. This standard SHF approach has
been shown to be very successful in describing the structure of finite nuclei,
especially the global properties such as binding energies and charge
radii~\cite{Cha97,Fri86,Klu09}. Instead of using directly the $9$ Skyrme interaction
parameters, we can express them explicitly in terms of $9$ macroscopic quantities,
i.e., $\rho _{0}$, $E_{0}(\rho_{0})$, the incompressibility $K_{0}$, the
isoscalar effective mass $m_{{\rm s},0}^{\ast }$, the isovector effective mass
$m_{{\rm v},0}^{\ast }$, $E_{\text{sym}}({\rho _{\rm r}})$, $L({\rho _{\rm r}})$, $G_{S}$,
and $G_{V}$. The $G_{S}$ and $G_{V}$ are respectively the gradient and
symmetry-gradient coefficients in the surface interaction part of the binding energies
for finite nuclei which is defined as
\begin{equation}
E_{\mathrm{grad}}=G_{S}(\nabla \rho )^{2}/(2{\rho )}-G_{V}\left[ \nabla
(\rho _{\rm n}-\rho _{\rm p})\right] ^{2}/(2{\rho )}.
\end{equation}
Then, by varying individually these macroscopic quantities within their known
ranges, one can examine more transparently the correlation of
nuclear matter properties with each individual macroscopic
quantity. Recently, this correlation analysis method has been
successfully applied to study the neutron skin~\cite{Che10,Zha13}
and giant monopole resonance of finite nuclei~\cite{Che12}, the higher-order bulk
characteristic parameters of ANM~\cite{Che11a}, and the relationship
between the nuclear matter symmetry energy and the symmetry energy
coefficient in nuclear mass formula~\cite{Che11}. In the present work, we
use the $9$ macroscopic quantities instead of using directly the $9$ Skyrme
interaction parameters. Especially, we study the symmetry energy
effects by varying the $L({\rho _{c}})$ value while keeping the other
macroscopic quantities unchanged.

\subsection{Method of Analyzing DM Direct Detection Experiments}
In this work, we use the standard method of analyzing DM direct detection
experiments~\cite{Lew96,Fran13}. The spin-independent differential event rate of
nuclear recoils with a recoil energy of $E_{\rm R}$, occurring in a detector
due to an elastic collision between a target nucleus of mass $m_{\rm N}^{\rm A}$
and a DM particle of mass $m_{\chi}$, can be expressed as
\begin{equation}
\frac{{\rm d} R^{\rm A}}{{\rm d} E_{\rm R}}
= \frac{\rho_{\chi} N_{\rm T}^{\rm A}}{m_{\rm N}^{\rm A} m_{\chi}}
  \int_{\vmin}^{v_{\rm E}+v_{\rm esc}} v_{\chi} f(\vdbf+\vebf) \frac{{\rm d} \sigma_{\rm p}}{{\rm d} E_{\rm R}}
  A^2 \left|F_{\rm A}(q)\right|^2 {\rm d}^{3}\vd,
\label{dRE}
\end{equation}
where A represents the mass number of the target nucleus and $N_{\rm T}^{\rm A}$ is
the number of target nucleus per unit mass in the detector, $\rho_{\chi}$ is the local
halo DM density, $f(v_{\chi})$ is the local DM velocity distribution evaluated in the
Galactic rest frame with $v_{\chi}=\left|\vdbf\right|$, $\vebf$ is Earth velocity
in the Galactic rest frame, and $v_{\rm esc}$ is the Galactic escape velocity~\cite{Gel01,Scho10}.
$\frac{{\rm d} \sigma_{\rm p}}{{\rm d} E_{\rm R}}$ and $F_{\rm A}(q)$ denote, respectively,
the spin-independent differential DM-proton scattering cross-section and the effective
form factors, and they will be detailed below. The $q=(2m_{\rm N}^{\rm A}E_{\rm R})^{1/2}$
is the momentum transfer between the DM particle and the struck nucleus.
The lower limit $\vmin$ of the integration in Eq.~(\ref{dRE}) corresponds to the smallest
DM velocity that can give a recoil energy of $E_{\rm R}$, i.e.,
\begin{equation}
\vmin(E_{\rm R})
= \sqrt{\frac{m_{\rm N}^{\rm A} E_{\rm R}}{2 \mu_{\rm A}^2}},
\label{vmin}
\end{equation}
where $\mu_{\rm A} = m_{\rm N}^{\rm A} m_{\chi} / (m_{\rm N}^{\rm A}+m_{\chi})$ is
the DM-nucleus reduced mass.
The predicted normalized recoil spectrum can thus be expressed as
\begin{equation}
f_{\rm s}(E_{\rm R}) =
\frac{\sum_{\rm A} \eta_{\rm A} \frac{{\rm d}R^{\rm A}}{{\rm d}E_{\rm R}}}
{\sum_{\rm A} \eta_{\rm A} \int \frac{{\rm d}R^{\rm A}}{{\rm d}E_{\rm R}} {\rm d}E_{\rm R}} \, ,
\label{spec}
\end{equation}
where $\eta_{\rm A}$ is the natural abundance of the isotope with mass number A for the
target element.

The total event rate $R_{\rm ex}$ expected in an energy range $[E_{1},E_{2}]$ in a detector
consisting of compound targets with finite detector energy resolution, usually characterized
by the unit of one event $kg^{-1} \, d^{-1}$, can be written as~\cite{Fran13}
\begin{equation}
R_{\rm ex} =
 \sum_{\rm A}  \eta_{\rm A} \int  {\rm d}E_{\rm R}
 \epsilon(E_{\rm R})  {\rm Res}(E_{\rm R},E_1,E_2)
 \frac{{\rm d}R^{\rm A}}{{\rm d}E_{\rm R}},
\label{tER}
\end{equation}
where $\epsilon(E_{\rm R})$ and ${\rm Res}(E_{\rm R},E_1,E_2)$ are the detector acceptance and
the detector response function, respectively. The $E_{\rm R}$ integration region is determined
by the cuts in the experiments.

It is convenient, and usually adequate, to describe the matter distributions
of a finite nucleus by a form factor, $F$, in analyzing DM direct detection experiments~\cite{Lew96}.
In the first-order Born approximation, the form factors of spherical nuclei are the
Fourier transforms of the proton and neutron density distributions, i.e.,
\begin{equation}
F_{\rm A}^{\rm{p,n}}(q) =
 \int  \rho_{\rm A}^{\rm{p,n}}(\boldsymbol{\mathit{r}})  e^{i\boldsymbol{\mathit{q}}\cdot\boldsymbol{\mathit{r}}}
 {\rm d}^{3}r
 =  \frac{4\pi}{q}  \int_{0}^{\infty}  r  \sin(qr)  \rho_{\rm A}^{\rm{p,n}}(r)
 {\rm d}r,
\label{ff}
\end{equation}
where the index p (n) denotes protons (neutrons). The density distributions,
$\rho_{\rm A}^{\rm{p,n}}$, for spherical nuclei can be obtained by using a
mean-field approach~\cite{RS80}, e.g., the SHF approach as we discussed above.

We further define the so-called effective form factors in Eq.~(\ref{dRE}) in terms of
$F_{\rm A}^{\rm{p,n}}$ as
\begin{equation}
\left | F_{\rm A}(q) \right|^{2}
= \frac{1}{A^2} \left | Z F_{\rm A}^{\rm p}(q) + g_{\rm np} N F_{\rm A}^{\rm n}(q) \right|^{2},
\label{eff}
\end{equation}
where Z (N) is the proton (neutron) number of the target nucleus, A=Z+N, and
$g_{\rm np}=f_{\rm n}/f_{\rm p}$ is the isospin-violation factor with $f_{\rm n}$
and $f_{\rm p}$ denoting the effective coupling of DM to neutrons and protons,
respectively. The spin-independent differential DM-proton scattering cross-section
can then be expressed as~\cite{Fran13}
\begin{equation}
\frac{{\rm d} \sigma_{\rm p}}{{\rm d} E_{\rm R}}
= \frac{m_{\rm N}^{\rm A} \sigma_{\rm p}}{2 \mu_{\rm p \chi}^2 v_{\chi}^2},
\label{dcs}
\end{equation}
where $\mu_{\rm p \chi}$ is the reduced DM-proton mass, and $\sigma_{\rm p}$
is the elastic DM-proton cross-section at zero momentum transfer ($q=0$).
Integrating the differential event rate in Eq. (\ref{dRE}) with respect to
$E_{\rm R}$, i.e., Eq.~(\ref{tER}), one can relate the expected number
of scattering events in a direct detection experiment with the DM-proton
cross-section $\sigma_{\rm p}$. Furthermore, by adopting, e.g., a profile
likelihood analysis proposed by the XENON100 collaboration~\cite{Profile11} or
the maximum gap method introduced by Yellin~\cite{Yel02} {\it et al.},
one can constrain the elastic DM-proton cross-section $\sigma_{\rm p}$ in
the $m_{\chi}$-$\sigma$ plane with specific confidence level.

However, in experiments usually reported is the DM-nucleon cross section
$\sigma_{\rm N}$, rather than $\sigma_{\rm p}$. $\sigma_{\rm N}$ is an
effective cross section by assuming that the DM particle couples with protons
and neutrons equally (i.e., $g_{\rm np}=1$) and the protons and neutrons have
identical form factor which is in practice commonly taken as the empirical
charge form factor parameterized by Helm~\cite{Helm56}. Obviously, the DM-proton
cross section $\sigma_{\rm p}$ is generally different from the DM-nucleon cross
section $\sigma_{\rm N}$, but they can be related to each other by the so-called
degradation factor $D_{\rm p}$ defined as~\cite{Feng13a}
\begin{equation}
D_{\rm p} \equiv \frac{\sigma_{\rm N}}{\sigma_{\rm p}}
= \frac{\sum_{\rm A} \eta_{\rm A} N_{\rm T}^{\rm A} A^2 \int
 \epsilon(E_{\rm R})  {\rm Res}(E_{\rm R},E_1,E_2)
 k(E_{\rm R}, m_{\chi}) \left | F_{\rm A}(E_{\rm R}) \right|^{2}{\rm d}E_{\rm R}}{
 \sum_{\rm A} \eta_{\rm A} N_{\rm T}^{\rm A} A^2 \int
 \epsilon(E_{\rm R})  {\rm Res}(E_{\rm R},E_1,E_2)
 k(E_{\rm R}, m_{\chi}) \left | F_{\rm A}^{\rm Helm}(E_{\rm R}) \right|^{2}{\rm d}E_{\rm R}},
\label{degra}
\end{equation}
with
\begin{equation}
k(E_{\rm R}, m_{\chi})
= \int_{v_{\rm min}(E_{\rm R},m_{\chi})}^{v_{\rm E}+v_{\rm esc}} f(\vdbf+\vebf) / v_{\chi} {\rm d}^{3}\vd \, ,
\label{maxinte}
\end{equation}
In Eq.~(\ref{degra}), $F_{\rm A}^{\rm Helm}$ is the Helm form factor which
is obtained by using a charge density with a Gaussian surface distribution
and has the following simple analytical expression
\begin{equation}
F_{\rm A}^{\rm Helm}(q)
 =  3  \frac{j_{1}(q r_{\rm N})}{q r_{\rm N}} \times
e^{-(qs)^2/2},
\label{helm}
\end{equation}
where $j_1$ is the first-order spherical Bessel function, $r_{\rm N}$ is an
effective nuclear radius and $s$ is a measure of the nuclear skin thickness.
For $r_{\rm N}$ and $s$, the following standard parametrization has been
used~\cite{Lew96}
\begin{eqnarray}
&r_{\rm N}^2&  =  c^2 + \frac{7}{3} \pi^2 a^2 - 5 s^2,  \\
&c&  \approx  1.23A^{1/3} - 0.60~{\rm fm},  \\
&a&  \approx  0.52 ~{\rm fm},  \\
&s&  \approx  0.9  ~{\rm fm}.
\end{eqnarray}
In Eq.~(\ref{degra}), it has been assumed that the variations of $g_{\rm np}$ and
form factors do not change the shape of recoil spectra $f_{\rm s}(E_{\rm R})$ in
Eq.~(\ref{spec}). In this case, varying $g_{\rm np}$ and form factors does not change
the event distributions but simply modifies the expected number of total events,
$R_{\rm ex}$ in Eq.~(\ref{tER}), for a given cross section. In general cases with
different $g_{\rm np}$ and nuclear form factors, since $R_{\rm ex}\propto \sigma_{\rm p}$,
we can simply rescale the corresponding cross sections according to the degradation
factor as $\sigma_{\rm p} = \sigma_{\rm N}/D_{\rm p}$ to keep the likelihood function
(the maximum gap) in a profile likelihood analysis (the maximum gap method)
unchanged. However, if the recoil spectrum changes significantly, the degradation factor in
Eq.~(\ref{degra}) should be modified according to the following identity
\begin{equation}
D_{\rm p} =
D_{\rm p}^{\rm 0} \times \frac{R_{\rm ex}(\sigma_{\rm N}^{\rm up})}{R_{\rm ex}(\sigma_{\rm p}^{\rm up})} \, ,
\label{degra2}
\end{equation}
where $D_{\rm p}^{\rm 0}$ denotes the degradation factor defined in Eq.~({\ref{degra}}),
$R_{\rm ex}(\sigma_{\rm p}^{\rm up})$ and $R_{\rm ex}(\sigma_{\rm N}^{\rm up})$ are
the number of expected events with $\sigma_{\rm p}^{\rm up}$ and $\sigma_{\rm N}^{\rm up}$
being the upper limits of the DM-proton and DM-nucleon cross sections with certain
confidence level. We estimate $R_{\rm ex}(\sigma_{\rm p}^{\rm up})$ and
$R_{\rm ex}(\sigma_{\rm N}^{\rm up})$ by using the ``maximum gap method''~\cite{Yel02}
in this work. In this way, one can thus study the isospin-violating effects and the
form factor effects simultaneously in analyzing the DM signals.

\section{Results and discussions}
\label{sec:sym}
%

\subsection{Symmetry energy effects on nuclear form factors}

As mentioned above, the neutron skin thickness of heavy nuclei is uniquely determined
by the symmetry energy density slope parameter $L(\rho_{\rm c})$ at a subsaturation
cross density $\rho_{\rm c} \approx 0.11 \, {\rm fm}^{-3}$~\cite{Zha13}. Therefore,
the $L(\rho_{\rm c})$ parameter controls the relative difference between the neutron
and proton distributions (and thus the neutron and proton form factors) in the nuclei.
In the present work, we investigate the form factor effects by varying the $L(\rho_{\rm c})$
parameter in the SHF calculations to fit the model-independent result of
$\Delta r_{\rm{np}}= 0.33_{-0.18}^{+0.16} \,\, {\rm fm}$ for $^{208}$Pb from the recent
PREX experiment at JLab.

\begin{figure}[tbp!]
\includegraphics[width=8.5cm]{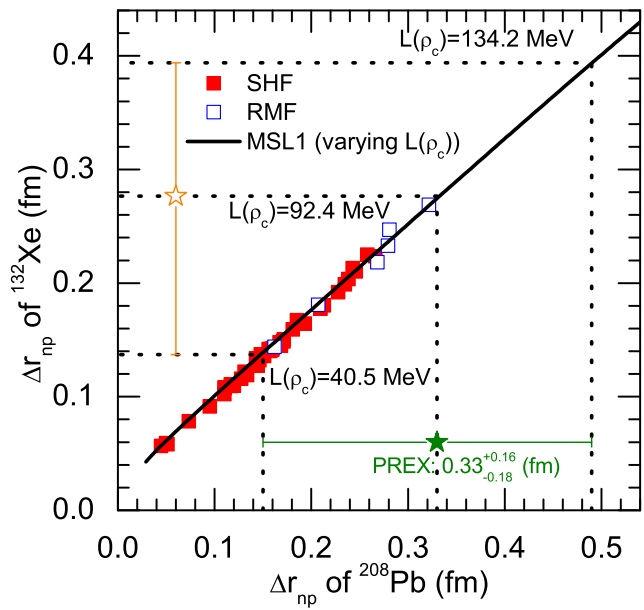}
\caption{(Color online) Neutron skin thickness for $^{132}$Xe vs. that for
$^{208}$Pb predicted by mean-field models with different interactions. Constraints
set by PREX measurement~\cite{PREX12} are also shown.}
\label{nsts}
\end{figure}

Shown in Fig.~\ref{nsts} is the $\Delta r_{\rm{np}}$ of $^{132}{\rm Xe}$ as a function
of that of $^{208}{\rm Pb}$ in SHF calculations with the Skyrme interaction MSL1~\cite{Zha13}
by varying $L(\rho_{\rm c})$ while keeping the other 8 macroscopic quantities and the
spin-orbit coupling constant $W_{0}$ fixed at their default values in the MSL1
interaction, namely, $\rho _{0} = 0.1586$ fm$^{-3}$, $E_{0}(\rho_{0}) = -15.998$ MeV,
the incompressibility $K_{0} = 235.12$ MeV, the isoscalar effective mass
$m_{s,0}^{\ast } = 0.806m$, the isovector effective mass $m_{v,0}^{\ast } = 0.706m$,
$E_{\text{sym}}({\rho _{c}}) = 26.67$ MeV, $G_{S} = 126.69$ MeV$\cdot$fm$^5$,
$G_{V} = 68.74$ MeV$\cdot$fm$^5$, and $W_{0} = 113.62$ MeV$\cdot$fm$^5$. The MSL1
interaction has been obtained by fitting a number of experimental data of finite
nuclei, including the binding energy, the charge rms radius, the neutron
$3p_{1/2}-3p_{3/2}$ energy level splitting in $^{208}$Pb, isotope binding energy
difference, and neutron skin data of Sn isotopes. For comparison, we also include in
Fig.~\ref{nsts} the results from SHF calculations with $43$ other Skyrme interactions
(BSk1, BSk4, BSk5, BSk7, BSk10, BSk14, BS15, Dutta, E-fit, Esigma-fit, Gsigma-fit,
KDE, KDE0, MSL0, RATP, Rsigma-fit, SGI, SGII, SK255, SK272, SKa, SkI1, SkI2, SkI5,
SKM, SkMP, SKM*, SKT1, SKT4, SKT5, SKT6, Skz0, Skz1, Skz2, Skz3, Skz4, Skz-1, SLy4,
SLy5, SLy9, Z-fit, Zsigma-fit, and ZsigmaS-fit) and RMF calculations with $6$ different
interaction parameter sets (FSU, IUFSU, TM1, PK1, NL3, and NL1). The references of
these Skyrme and RMF interactions can be found in Refs.~\cite{ChenR12,ChenLW07,Fat10}.
We have selected these interactions in such a way that their $L(\rho_c)$ values scatter
in large region and are not close to each other. One can see clearly a nice
model-independent linear correlation between the neutron skin thicknesses of
$^{208}{\rm Pb}$ and $^{132}{\rm Xe}$ within the non-relativistic and relativistic
models with different interactions. In addition, although there are few interactions
that predict a $\Delta r_{\rm{np}}$ of $^{208}{\rm Pb}$ larger than $0.33$ fm
measured by the model-independent PREX experiment, one can easily obtain a large
$\Delta r_{\rm{np}}$ of $^{208}{\rm Pb}$ by increasing the $L(\rho_{\rm c})$ value
in the MSL1 interaction. In particular, we find a value of $L(\rho_{\rm c})=92.4_{-51.9}^{+41.8}$
MeV in the MSL1 interaction predicts $\Delta r_{\rm{np}}=0.33_{-0.18}^{+0.16}$ fm
for $^{208}$Pb and $\Delta r_{\rm{np}}=0.28_{-0.14}^{+0.12}$ fm for $^{132}{\rm Xe}$.
Therefore, we use $L(\rho_{\rm c}) = 40.5$ MeV, $92.4$ MeV and $134.2$ MeV in the MSL1
interaction, denoted as Lc40, Lc92 and Lc134, respectively, to study the symmetry
energy and form factor effects in the present work. We would like to point out that
the large neutron skin thickness for $^{208}$Pb measured by the PREX experiment can
also be explained within the RMF model~\cite{Fat13}.

\begin{figure}[tbp!]
\includegraphics[width=13.5cm]{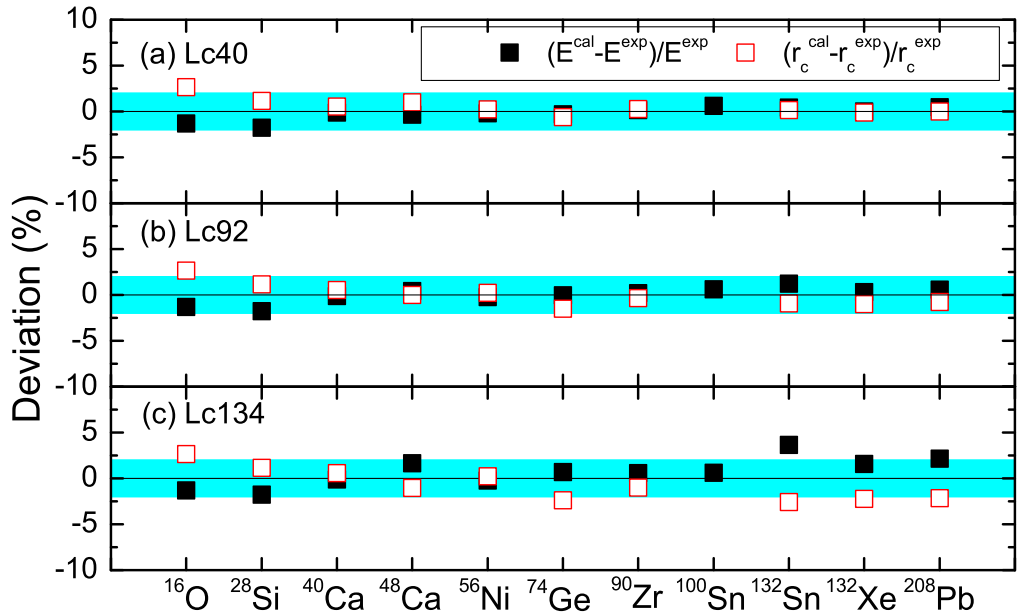}
\caption{(Color online) Deviations of the binding energies (solid squares) and
charge rms radii (open squares) of a number of nuclei obtained from SHF with
Lc40, Lc92 and Lc134 from those measured in experiments. The bands indicate
a deviation within $\pm 2\%$.}
\label{ebrc}
\end{figure}

To test the Skyrme interaction parameter sets Lc40, Lc92, and Lc134, we calculate
the binding energies and charge rms radii for a number of closed-shell or
semi-closed-shell nuclei, i.e., $^{16}$O, $^{40}$Ca, $^{48}$Ca, $^{56}$Ni, $^{90}$Zr,
$^{100}$Sn, $^{132}$Sn, and $^{208}$Pb, as well as the nuclei $^{28}$Si, $^{74}$Ge,
and $^{132}$Xe. The elements Si, Ge and Xe are widely used as targets in the DM
direct detection experiments. For Si and Ge, their natural abundances are mostly
dominated by their isotopes of $^{28}$Si and $^{74}$Ge, respectively. For Xe, however,
it has several isotopes with comparable natural abundances and here we just select
one of them, i.e. $^{132}$Xe, as an example to show its properties. Fig.~\ref{ebrc}
shows the relative deviation of the binding energies and charge rms radii of these
nuclei from those measured in experiments~\cite{Wang12,Angeli04,Blanc05}. It is seen
that the interactions Lc40 and Lc92 can describe the experimental data very well
(the deviations are within about $\pm 1\%$) except for the light nucleus $^{16}$O
for which the deviation of charge rms radius is about $2.6\%$. It is interesting to
see that, the interaction Lc134, which predicts a very strong density dependence
of the symmetry energy at $\rho_c$ and gives a very large neutron skin thickness
of $\Delta r_{\rm{np}}=0.49$ fm for $^{208}{\rm Pb}$ (the upper limit of
the PREX measurement), still can give a reasonable description of the experimental
data (the deviations are within about $\pm 2\%$ as indicated by bands in
Fig.~\ref{ebrc}) except the nucleus $^{132}$Sn for which the deviation of
charge rms radius (binding energy) reaches about $-2.6\%$ ($3.7\%$) and for
the light nucleus $^{16}$O the deviation of charge rms radius is still about
$2.6\%$. These results are remarkable as Lc40, Lc92 and Lc134 are not obtained
from fitting measured binding energies and charge rms radii of finite nuclei
as in usual Skyrme parametrization. It should be pointed out that our main
motivation for introducing the interactions Lc40, Lc92 and Lc134 is not to construct
new Skyrme interaction parameter sets to describe data, but to use them as
references to study the form factor effects in the following.

\begin{figure}[tbp]
\includegraphics[width=13.5cm]{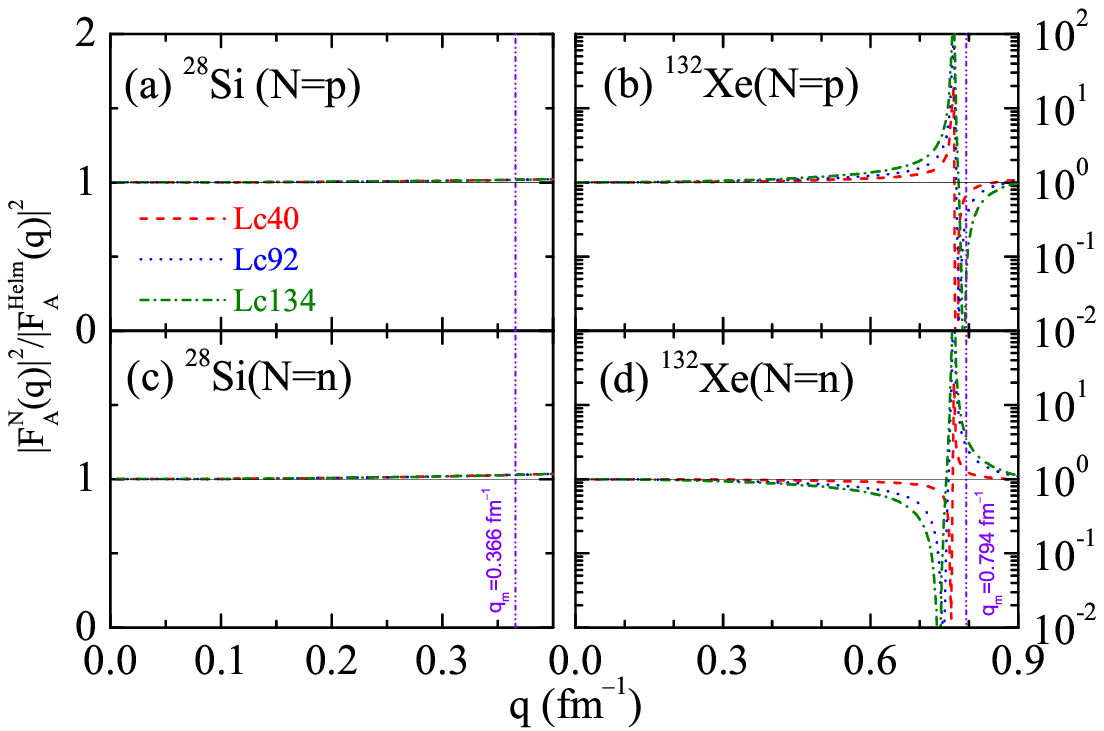}
\caption{(Color online) Proton (upper panels) and neutron (lower panels) form
factors divided by the corresponding empirical Helm form factors as functions
of momentum transfer $q$ for $^{28}$Si (left panels) and $^{132}$Xe (right panels)
obtained from SHF with Lc40, Lc92 and Lc134. The momentum transfer $q_m$,
corresponding to a value of $E_{\rm R}=100$ keV, is indicated by dash-dot-dotted
lines.}
\label{fnp}
\end{figure}

To investigate the symmetry energy effects on the proton and neutron form factors
in finite nuclei, we plot in Fig.~\ref{fnp} the proton (upper panels) and neutron
(lower panels) form factors as functions of momentum transfer $q$  for $^{28}$Si
(left panels) and $^{132}$Xe (right panels) from SHF calculations with Lc40, Lc92
and Lc134. For comparison, the results are divided by the corresponding
empirical Helm form factors. The proton and neutron form factors of other isotopes
or elements can be obtained from a generalized Helm-like empirical parametrization
given in the Appendix where the proton and neutron form factors are distinguished
and parameterized as functions of the neutron skin thickness of $^{208}$Pb by
fitting the results obtained from SHF calculations with Lc40, Lc92 and Lc134.
In addition, also indicated in Fig.~\ref{fnp} is the momentum transfer $q_m$
corresponding to a value of $E_{\rm R}=100$ keV which represents the
possible maximum cutoff among various experiments. However, we would like to
emphasize that the upper limits of the recoil energy $E_{\rm R}$ usually vary
significantly in different DM direct detection experiments, and here we just
select a relatively large recoil energy of $100$ keV as an example to see the
possible symmetry energy effects on the proton and neutron form factors in
different nuclei. In calculations of the scattering cross sections
in Eq.~(\ref{degra}) for each experiment, one must use the recoil energy region
given by each experiment. It is seen from Fig.~\ref{fnp} that, for the nucleus
$^{28}$Si with equal proton and neutron numbers, the symmetry energy effect is
tiny and the SHF calculations with Lc40, Lc92 and Lc134 predict almost the same
form factors for neutrons and protons, which are further in good agreement with
the empirical Helm form factor for $q \le q_m$. On the other hand, for the neutron-rich
nucleus $^{132}$Xe, one can see a clear symmetry energy effect on the form
factors, namely, a larger $L(\rho_c)$ shifts the neutron form factor to lower
$q$ values while shifts slightly the proton form factor to higher $q$ values with
the empirical Helm form factor in between, leading to an isospin splitting
between the proton and neutron form factors with respect to the $q$ value. These
features actually reflect the symmetry energy effects on the neutron skin
thickness, namely, increasing the value of $L(\rho_c)$ increases the rms
radius of neutrons and reduces slightly the rms radius of protons, and thus
leads to a larger neutron skin thickness as shown in Fig.~\ref{nsts}.

\begin{figure}[tbp]
\includegraphics[width=13.5cm]{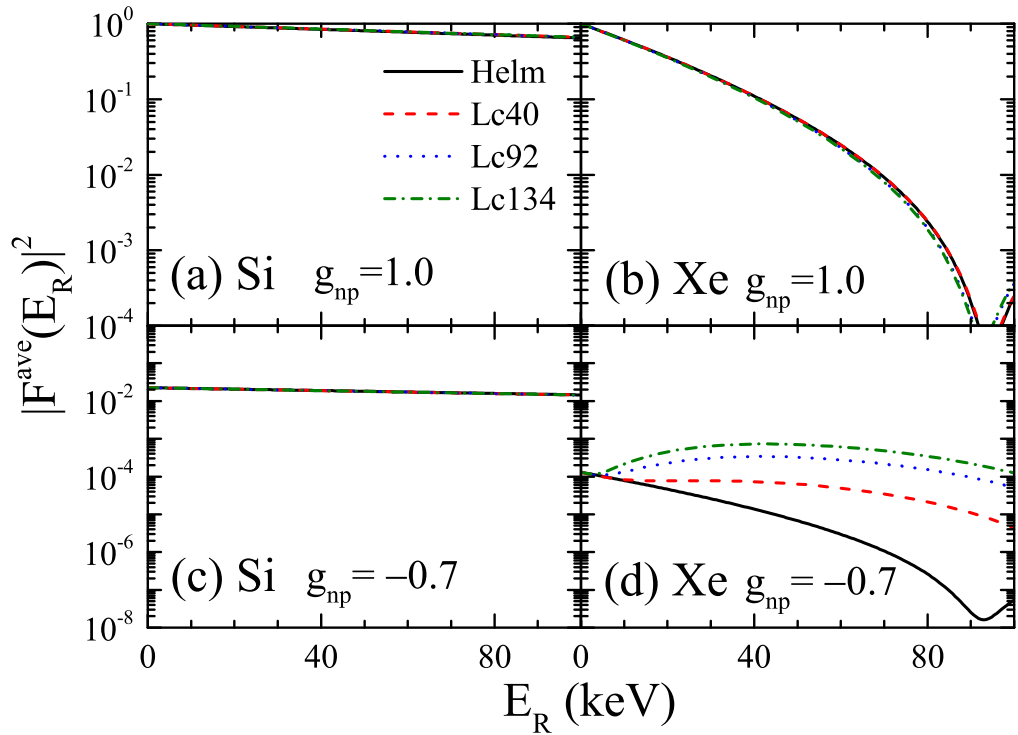}
\caption{(Color online) The averaged effective form factor as a function
of recoil energy $E_{\rm R}$ for Si (left panels) and Xe (right panels) with
$g_{\rm np} = 1.0$ (upper panels) and $g_{\rm np} = -0.7$ (lower panels)
obtained from SHF with Lc40, Lc92 and Lc134. The empirical Helm form
factor is also included for comparison.}
\label{ffs}
\end{figure}
Since the effective form factor (Eq. ({\ref{eff}})) should be used to analyze
the DM signals for general cases, it is expected that the isospin splitting
between the proton and neutron form factors would cause significant effects on the
effective form factor, especially for a negative $g_{\rm np}$ which could cause
strongly destructive interference of DM scattering with protons and neutrons.
On the other hand, the presence of different isotopes with comparable abundances
in some detector target elements (e.g., xenon) suggests that it is better to
study the averaged effective form factors defined as
\begin{equation}
\left | F^{\rm ave}(E_{\rm R}) \right|^{2}
= \frac{\sum_{\rm A} \eta_{\rm A} N_{\rm T}^{\rm A} A^{2} \left | F_{\rm A}(q(E_{\rm R})) \right|^{2}}{\sum_{\rm A} \eta_{\rm A} N_{\rm T}^{\rm A} A^{2}}.
\label{eff_tot}
\end{equation}
This is illustrated in Fig.~\ref{ffs} where the averaged effective form
factor is plotted as a function of $E_{\rm R}$ for Si (left panels) and Xe (right panels)
from SHF calculations with Lc40, Lc92 and Lc134. The results from the empirical
Helm form factor with $F_{\rm A}^{\rm{p,n}}(q)=F_{\rm A}^{\rm Helm}(q)$
are also included for comparison. In Fig.~\ref{ffs}, two cases for the DM
are considered: one is for the  standard isospin-invariant DM with $g_{\rm np} = 1.0$
(upper panels) and the other is for the IVDM with $g_{\rm np} = -0.7$ (lower panels).
It should be pointed out that $g_{\rm np}=-0.7$, firstly suggested by Feng
{\it et al.}~\cite{Feng11}, leads to nearly complete destructive interference
of the scattering amplitudes for DM-proton and DM-neutron collisions in Xe target
and gives maximum suppression of the sensitivity for Xe-based detectors if the
empirical Helm form factor is adopted. For such kind of IVDM, the confidence region
of CDMS-II(Si) and the XENON100 exclusion contours are consistent with each
other~\cite{Fran13}. As will be shown later, for IVDM with $g_{\rm np}=-0.7$, both
the results from recent LUX~\cite{LUX13} and the most recent SuperCDMS(Ge)~\cite{SCDMS14}
are also consistent with the confidence region of CDMS-II(Si). Theoretically, such
negative values of $g_{\rm np}$ can arise, e.g., in models with a new light
neutral gauge boson $Z'$~\cite{Fra11,Gao13}.

One can see from Fig.~\ref{ffs} that the negative $g_{\rm np}$ (i.e., $-0.7$)
indeed causes strongly destructive interference of DM scattering with protons and
neutrons for both Si and Xe targets and thus the averaged effective form factor with
$g_{\rm np}=-0.7$ is strongly suppressed compared to its value with $g_{\rm np}=1.0$.
Furthermore, it is interesting to see from Fig.~\ref{ffs} that for $g_{\rm np}=1.0$,
the averaged effective form factors of both Si and Xe targets from SHF calculations
with Lc40, Lc92 and Lc134 are in a very good agreement with the corresponding
results from the empirical Helm form factors, indicating that there are essentially
no symmetry energy effects. However, for $g_{\rm np}=-0.7$, one can see that the
averaged effective form factors of Xe target display very different behaviors for
different distributions from SHF calculations with Lc40, Lc92 and Lc134 as well as
that from the empirical charge distributions, although the averaged effective form
factors of Si target are essentially the same for these different distributions.
These features indicate that the symmetry energy effect is tiny for the averaged
effective form factor of Si target but is very strong for that of Xe target in the
case of $g_{\rm np}=-0.7$. This is due to the fact that for a negative $g_{\rm np}$,
the scattering amplitudes for DM-proton and DM-neutron collisions may interfere
destructively, and the DM particle is almost completely decoupled from the Xe isotopes,
i.e. $F(E_{\rm R}) \approx 0$ (see panel (d) of Fig.~\ref{ffs}), especially for
$g_{\rm np} \approx -0.7$~\cite{Feng11}. In this case, a small difference between
$F^{\rm{p}}_{\rm A}$ and $F^{\rm{n}}_{\rm A}$ may lead to a significant change for
the effective form factor. As a result, the form factor of Xe exhibits a very strong
symmetry energy effect for $g_{\rm np}=-0.7$, namely, a larger $L(\rho_{\rm c})$ value
leads to a larger form factor as shown in panel (d) of Fig.~\ref{ffs}. In particular,
one can see that the averaged effective form factors of Xe target from SHF calculations
with Lc40, Lc92 and Lc134 can be significantly larger than that from the empirical Helm
form factor  in a large range of recoil energy. This implies that the symmetry energy
with a larger $L(\rho_{\rm c})$ (and thus a larger neutron skin thickness) can give a
more significant enhancement for the effective DM-nucleus form factors in Eq.~({\ref{dRE}})
when the IVDM with $g_{\rm np}=-0.7$ is considered. Therefore, for $g_{\rm np}=-0.7$,
one can expect a significant enhancement for the sensitivity of the Xe-based detectors
to the DM-proton cross sections by taking into account the form factor effects due to
different symmetry energies compared with the corresponding result using the empirical
Helm form factor in analyzing the experimental data.

\subsection{Form factor effects on the extraction of DM-proton cross sections}
\begin{figure}[tbp]
\includegraphics[width=13.5cm]{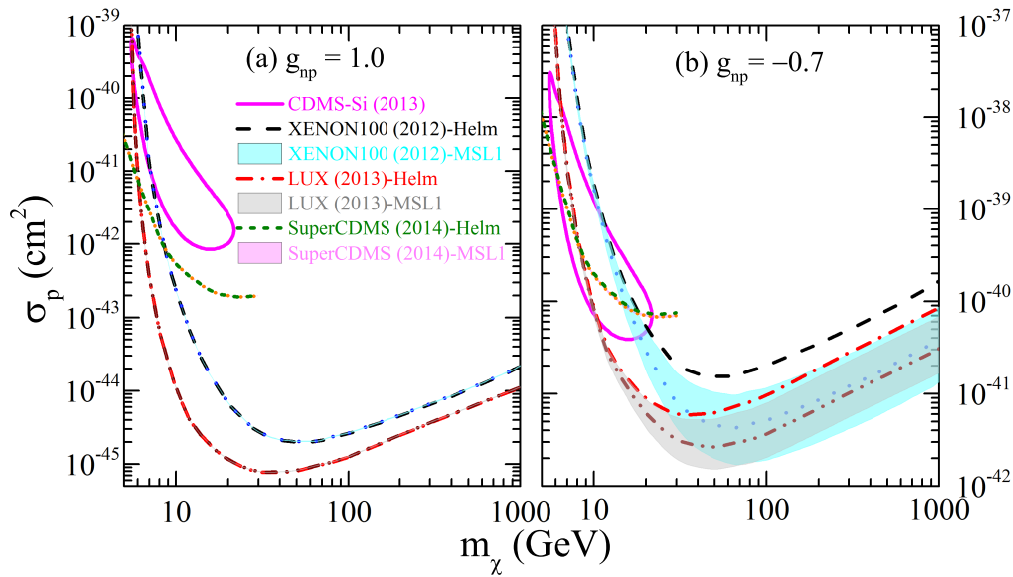}
\caption{(Color online) Results of $90 \%$ confidence level (C.L.) limits from
XENON100, LUX and SuperCDMS(Ge) together with $90 \%$ C.L. favored regions
from CDMS-II(Si) in $m_{\chi}$-$\sigma_{\rm p}$ plane for $g_{\rm np}=1.0$ (left panel)
and $g_{\rm np}=-0.7$ (right panel) with form factors from the empirical
charge distributions (Helm) and SHF calculations with MSL1 by varying $L(\rho_c)$ from
$40.5$ MeV to $134.2$ MeV.}
\label{css}
\end{figure}

To see the effects of the form factor and the isospin-violation factor
$g_{\rm np}$ on the extraction of elastic spin-independent DM-proton cross sections
$\sigma_{\rm p}$ in DM direct detection experiments, we show in Fig.~\ref{css}
the results of the $90 \%$ confidence level (C.L.) limits from
XENON100~\cite{XENON100v12}, LUX~\cite{LUX13} and SuperCDMS(Ge)~\cite{SCDMS14}
along with the $90 \%$ C.L. favored regions from CDMS-II(Si)~\cite{CDMS13v1}
in $m_{\chi}$-$\sigma_{\rm p}$ plane for $g_{\rm np}=1.0$ (left panel)
and $g_{\rm np}=-0.7$ (right panel) with form factors from the empirical
charge distributions and SHF calculations with MSL1 by varying $L(\rho_c)$ from
$40.5$ MeV to $134.2$ MeV. For the results from SuperCDMS(Ge), only the mass
region less than $30$ GeV is shown since no data are available for
$m_{\chi} > 30$ GeV~\cite{SCDMS14}. In these analyses, we use the standard
astrophysical parameters in the Standard Halo Model, namely, a Maxwell-Boltzmann
distribution for $f(v)$ with $v_0=220 \, {\rm km}/{\rm s}$ and Galactic escape
velocity $v_{\rm esc}=544 \, {\rm km}/{\rm s}$, and a DM density of
$\rho_{\chi}=0.3 \, {\rm GeV}/{\rm cm}^3$~\cite{Smith07,Pif14}.

In Fig.~\ref{css}, the form factor effects due to the symmetry energy and the
effects from $g_{\rm np}$ are considered through dividing the experimental results
by the degradation factor $D_{\rm p}$ as discussed earlier. In particular, for
XENON100, we assume that the energy resolution is dominated by Poisson fluctuations
in the number of photoelectrons (PE) as in Ref.~\cite{Fran13}. We use S1 efficiency
obtained from Fig. 1 of Ref.~\cite{XENON100v12} and the scintillation efficiency
$\mathcal{L}_{\rm eff}$ from Fig. 1 of Ref.~\cite{Profile11}, respectively, and
integrate the differential rate over S1 from $3$ to $30$ PE to get the total rate.
For the LUX analysis which is similar with XENON100's, the signal region is limited
within $2$ to $30$ PE and the efficiencies are taken from Fig. 1 of Ref.~\cite{LUX13}.
For CDMS-II(Si), we use the acceptance from Fig. 1 of Ref.~\cite{CDMS13v2}, and choose
an energy interval between $7 \, {\rm keV}$ and $100 \, {\rm keV}$ in the analysis.
For SuperCDMS(Ge), with the efficiency from Fig. 1 of Ref.~\cite{SCDMS14}, we use the
signal region in the range $1.6 - 10 \, {\rm keV}$ in the calculations as in
Ref.~\cite{SCDMS14}. Furthermore, since XENON100 reported $2$ candidate events in
the low recoil energy region, we calculate the degradation factor according to
Eq.~(\ref{degra2}) instead of Eq.~(\ref{degra}) to take into account the effects
due to shape changing of the recoil spectrum as we will show later.
Similarly, for LUX, we follow the analysis in Ref.~\cite{Del14} and calculate
the degradation factor according to Eq.~(\ref{degra2}) by using one observed event
(with ${\rm S1} = 3.1$ PE). For CDMS-II(Si) and SuperCDMS(Ge) that focus on the
light DM, we have checked that the shape changing of their recoil spectra can be
negligible and Eq.~(\ref{degra}) is thus applied to evaluate the degradation factor.

For the isospin-invariant DM with $g_{\rm np}=1.0$, one can see from Fig.~\ref{css}
that the CDMS-II(Si) results are in some tension with the upper limit placed by the
XENON100 experiment, and they become in even strong disagreement with the upper
limits set by the recent LUX~\cite{LUX13} experiment and the more recent
SuperCDMS(Ge)~\cite{SCDMS14} experiment. However, for the IVDM with $g_{\rm np}=-0.7$,
one can see that the tension between CDMS-II(Si) and XENON100 is essentially
disappeared as found by Frandsen~{\it et al.}~\cite{Fran13}. Furthermore, it is
remarkable to see from the right panel of Fig.~\ref{css} that the disagreement of
the CDMS-II(Si) results with LUX and SuperCDMS(Ge) can also be largely ameliorated
within the framework of IVDM with $g_{\rm np}=-0.7$ (We note that a similar
conclusion is also obtained in Ref.~\cite{Hamaguchi14}). Therefore, these results
indicate that IVDM indeed provides a very promising mechanism to reconcile
the tension among various experiments. It should be mentioned here that although
$g_{\rm np}=-0.7$ leads to a maximum suppression for the sensitivity of Xe-based
detector for the empirical Helm form factor, it does not for the more realistic
SHF form factors as will be shown later.

Furthermore, it is interesting to see that, although there are essentially no
form factor effects on the extracted spin-independent DM-proton cross
sections $\sigma_{\rm p}$ in different experiments in the case of
isospin-invariant DM with $g_{\rm np}=1.0$, the form factor effects can
significantly affect the extraction of $\sigma_{\rm p}$ for the Xe-based
experiments (XENON100 and LUX) in the case of IVDM with $g_{\rm np}=-0.7$.
Therefore, our results indicate that, for isospin-invariant DM, the widely
used empirical Helm form factor is well grounded and this is consistent with
our previous discussions about the form factor effects. On the other hand, for IVDM
with $g_{\rm np}=-0.7$, using the form factors obtained from SHF calculations in
MSL1 with varied $L(\rho_c)$ values generally increases the sensitivity of
the Xe-based detectors (XENON100 and LUX) compared with using the empirical
Helm form factor, and a larger $L(\rho_c)$ value, which leads to a larger
neutron skin thickness in Xe isotopes, generally leads to a stronger
sensitivity of the Xe-based detectors. This form factor effect becomes
more pronounced with the increment of the DM mass $m_{\chi}$. While the
form factor effect is about $20\%$ for the DM mass of $m_{\chi} \approx 8$ GeV,
it can become very significant for the DM with mass above a few tens GeV.
For $m_{\chi} = 20$ GeV, which roughly corresponds to the DM mass upper
limit of CDMS-II(Si) bounds, the sensitivities of XENON100 and LUX can be
enhanced by a factor of $3$ using the form factor with $L(\rho_{\rm c})=134.2$ MeV
compared with that using the empirical Helm form factor. In particular,
the sensitivities can have a factor of more than $10$ improvement for
$m_{\chi} \geq 80$ GeV where many constraints for supersymmetric WIMPs have
been put by the data from LHC~\cite{SUSYDM}. On the other hand, the relative
variation between the results extracted from the Helm form factor and
the SHF calculations is small ($<7 \%$) for SuperCDMS(Ge) and it can be
negligible ($<0.5 \%$) for CDMS-II(Si) for which the results from the
SHF calculations are not shown in Fig.~\ref{css}.

\begin{figure}[tbp]
\includegraphics[width=14cm]{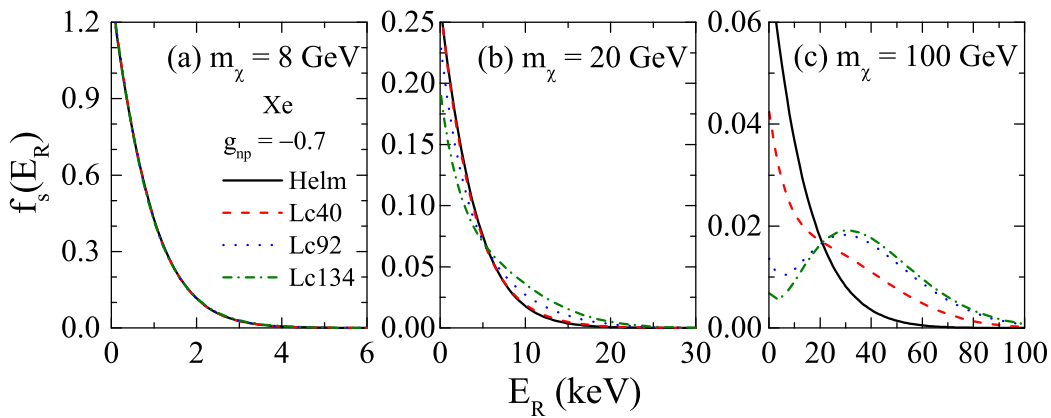}
\caption{(Color online) The normalized DM-xenon recoil spectrum $f_{\rm s}(E_{\rm R})$
with $g_{\rm np}=-0.7$ for $m_{\chi}=8 \, {\rm GeV}$ (panel (a)), $m_{\chi}=20 \, {\rm GeV}$
(panel (b)) and $m_{\chi}=100 \, {\rm GeV}$ (panel (c)), respectively. Cases for different
form factors from the empirical Helm charge distribution (black solid lines) and SHF
calculations with MSL1 by varying $L(\rho_c)$ from $40.5$ MeV to $134.2$ MeV are shown in
the figure. For comparison, spectra calculated with Helm form factors for isospin-invariant
cases ($g_{\rm np} = 1$) are also shown.}
\label{spe}
\end{figure}

The DM mass dependence of the form factor effects observed in XENON100
and LUX for $g_{\rm np}=-0.7$ can be understood as a result of the shape changing
of the expected recoil spectra for varied $m_{\chi}$ due to the enhancement effects
of the form factors when we take the symmetry energy effects into account
as discussed in the previous subsection. Shown in Fig.~{\ref{spe}} is the normalized
DM-xenon recoil spectrum $f_{\rm s}(E_{\rm R})$ for $m_{\chi}=8\, {\rm GeV}$, $20 \,
{\rm GeV}$ and $100 \, {\rm GeV}$ with $g_{\rm np}=-0.7$ and various form factors
from SHF calculations with Lc40, Lc92 and Lc134 as well as the empirical Helm form
factor. We note that for $g_{\rm np}=1.0$, the $f_{\rm s}(E_{\rm R})$ essentially does
not depend on the variation of the form factors and the results are almost the same
as that from the empirical Helm form factor with $g_{\rm np}=-0.7$. It is very
interesting to see from Fig.~{\ref{spe}} that for the case with $g_{\rm np}=-0.7$,
although the form factor effect essentially does not change or only slightly modifies
the predicted normalized DM-xenon recoil spectra for light DM (see, e.g., Figs.~\ref{spe}
(a) and (b)), it can drastically change the shape of the recoil spectra for heavy DM
(see, e.g., Fig.~\ref{spe} (c)). For both XENON100 and LUX, we have considered this
spectrum shape changing effect in Eq.~(\ref{degra2}) by multiplying Eq.~(\ref{degra})
by another term $R_{\rm ex}(\sigma_{\rm N}^{\rm up})/R_{\rm ex}(\sigma_{\rm p}^{\rm up})$
for the degradation factor, which will result in, e.g., $\sim 14 \%$ ($\sim 7 \%$)
enhancement of the upper limit of $\sigma_{\rm p}$ from XENON100 (LUX) relative to the
results obtained from Eq.~(\ref{degra}) for the Lc134 case with $m_{\chi} = 100 \, {\rm GeV}$.
More specifically, one can see from Fig.~{\ref{spe}} that for heavier DM, the expected
DM-xenon recoil spectra become harder and thus the effective form factors are
typically probed at higher momentum transfer, and this will significantly enhance the
effects caused by different form factors (see, e.g., Fig.~\ref{ffs} (d)) adopted for
the $E_{\rm R}$ integration in the numerator of r.h.s of Eq.~(\ref{degra}), leading
to the observed DM mass dependence of the form factor effects in XENON100 and LUX for
$g_{\rm np}=-0.7$.

It is also very interesting to see that for the case with $g_{\rm np}=-0.7$, the bounds
of XENON100 and LUX overlap with each other in the mass region of $m_{\chi} \ge 70$ GeV
although the LUX gives much more stringent limit on the spin-independent elastic DM-proton
scattering cross section than XENON100 in the lower mass region as shown in the right
panel of Fig.~\ref{css}. This can also be explained as a result of the shape changing of
the expected recoil spectrum for varied $m_{\chi}$ as shown in Fig.~{\ref{spe}}.
Particularly, Fig.~\ref{spe} suggests that a significant fraction of the
events is expected at larger recoil energies for heavier DM, and this is especially
the case for the IVDM with $g_{\rm np}=-0.7$ and SHF form factors with Lc92 and
Lc134 where one can see the non-standard recoil spectra peaked around
$E_{\rm R} = 30$ keV. Since LUX adopts a relatively small energy range, i.e.
$3.0-22.1$ keV~\cite{LUX13}, compared with XENON100 where $6.6-43.3$ keV is
adopted~\cite{XENON100v12}, it thus exhibits relatively weaker sensitivity than
XENON100 for heavier DM. As a matter of fact, even for the isospin-invariant
case where the form factor effects are negligible as shown in Fig.~{\ref{css}}~(a),
the sensitivity of LUX becomes relatively weaker with the increment of $m_{\chi}$
compared with that of XENON100. It should be mentioned that the small form factor
effect observed for SuperCDMS(Ge) in Fig.~{\ref{css}}~(b) is also partially due to
the smaller threshold energy cut, i.e., $1.6-10.0$ keV~\cite{SCDMS14}. Our results
imply that the non-standard recoil spectra shown in Fig.~\ref{spe}~(c) may provide
an extremely important experimental evidence for non-standard dark matter
interactions and they also give valuable information on the direct detection of IVDM.

\subsection{Form factor effects with various isospin-violation factor $g_{\rm np}$}
\begin{figure}[tbp]
\includegraphics[width=13.5cm]{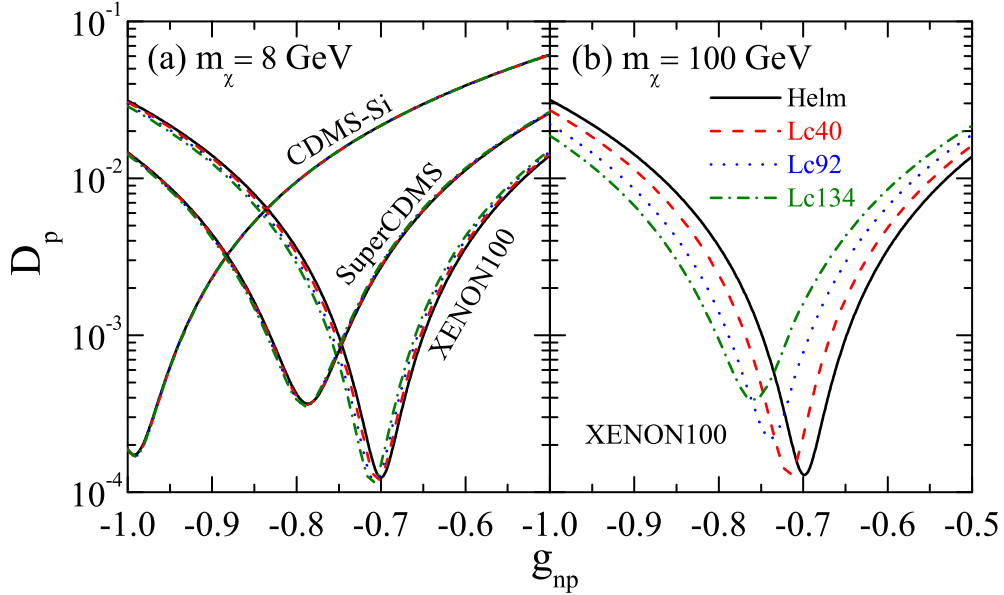}
\caption{(Color online) Degradation factor $D_{\rm p}$ as a function of the
isospin-violation factor $g_{\rm np}$ obtained by using the
empirical Helm form factors as well as the form factors from SHF calculations with
Lc40, Lc92 and Lc134 for $m_{\chi}=8$ GeV (panel (a)) and $m_{\chi}=100$ GeV
(panel (b)). For $m_{\chi}=8$ GeV, the results of CDMS-II(Si), SuperCDMS(Ge) and
XENON100 are shown while only the results of XENON100 are shown for $m_{\chi}=100$ GeV.}
\label{xepho}
\end{figure}

In above calculations for the case of IVDM, we have fixed the isospin-violation
factor at $g_{\rm np}=-0.7$. As mentioned earlier, for the specific IVDM with
$g_{\rm np}=-0.7$, the sensitivity of Xe-based detectors is maximumly suppressed
when the empirical Helm form factor is used and the tension among various experiments
using different target elements has been shown to be largely ameliorated.
For more realistic SHF form factors, it is interesting to see how the form factor
effects change with more general values of the isospin-violation factor $g_{\rm np}$.
Following the method by Feng~{\it et al.}~\cite{Feng13a}, we study in the following
how the form factor effects change for various isospin-violation factor $g_{\rm np}$.
Shown in Fig.~\ref{xepho} is the degradation factor $D_{\rm p}$ as a function of
the isospin-violation factor $g_{\rm np}$ obtained by using the empirical Helm form
factors as well as the form factors from SHF calculations with Lc40, Lc92 and Lc134
for $m_{\chi}=8$ GeV (panel (a)) and $m_{\chi}=100$ GeV (panel (b)). Since both XENON100
and LUX are Xe-based experiments and their results have similar dependence on the $g_{\rm np}$
and the symmetry energy, we thus only consider CDMS-II(Si), SuperCDMS(Ge) and XENON100 in
Fig.~\ref{xepho}~(a) for $m_{\chi}=8$ GeV. In addition, for $m_{\chi}=100$ GeV, the form
factor effect is negligible for CDMS-II(Si) and relatively small for SuperCDMS(Ge)
which adopts a small threshold energy cut and has been designed especially for light
DM searching, we thus only show the results of XENON100 (panel (b)).

It is seen from Fig.~\ref{xepho} that generally the sensitivity of each detector is
significantly suppressed for a negative $g_{\rm np}$ within the range of $-1.0<g_{\rm np}<-0.5$
due to destructive interference between DM-neutron and DM-proton scattering amplitudes as
pointed out in Ref.~\cite{Feng11,Feng13a}. However, the $D_{\rm p}$ reaches its minimum value
(where the detector is most insensitive) at different values of $g_{\rm np}$ for different
detectors due to the different compositions and distributions of neutrons and protons in each
target element of the detector. In particular, one can see that the corresponding value
of $g_{\rm np}$ which leads to a minimum value of $D_{\rm p}$, denoted as $g_{\rm np}^{\rm min}$,
is about $-0.993$, $-0.787$ and $-0.700$ for CDMS-II(Si), SuperCDMS(Ge) and XENON100,
respectively, for the case of $m_{\chi}=8$ GeV when the empirical Helm form factor is used.
For $m_{\chi}=8$ GeV, the form factor effect is very small for CDMS-II(Si) and the largest
relative variation with respective to the results from the empirical Helm form factor
when $L(\rho_c)$ varies from $40.2$ MeV to $134.2$ MeV is $1.26\%$ at $g_{\rm np}=-0.997$.
The largest relative variation due to the form factor effect is $6.22\%$ at $g_{\rm np}=-0.818$
for SuperCDMS(Ge), and it becomes $47.11\%$ at $g_{\rm np}=-0.686$ for XENON100. As a matter
of fact, a clear shift of curves for the Xe-based detectors calculated with different form
factors has been already seen even for such a small DM mass in Fig.~\ref{xepho}~(a). These
features indicate that the form factor effect can be already pronounced for XENON100 even for
a small mass DM with $m_{\chi}=8$ GeV. For XENON100, one can see from Fig.~\ref{xepho}(b) that
the form factor effect becomes very strong for $m_{\chi}=100$ GeV. In particular, a larger
$L(\rho_c)$ value gives significantly stronger sensitivity of the detectors and the largest
relative variation due to the form factor effect is $952.30\%$ (almost one order) at
$g_{\rm np}=-0.695$.

Furthermore, one can see from Fig.~\ref{xepho} that the value of $g_{\rm np}^{\rm min}$ also
depends significantly on the form factor, and it is $-0.716$, $-0.737$ and  $-0.754$ for the
SHF form factor with $L(\rho_c)=40.2$, $92.4$ and $134.2$ MeV, respectively, smaller than
$g_{\rm np}=-0.7$ for the empirical Helm form factor. In addition, it is very interesting
to see that, while using the SHF form factors enhances the sensitivity of the Xe-based
detector relative to the case with Helm form factor for $g_{\rm np} \ge -0.7$, it suppresses
the sensitivity for $g_{\rm np} < -0.7$. These features indicate that the form factor effects
can either enhance or suppress the sensitivity of the detector relative to the Helm's case
depending on the specific value adopted for $g_{\rm np}$. It is also very interesting to see
that one cannot find a $g_{\rm np}$ value such that $D_{\rm p} \rightarrow 0$, leading to
zero sensitivity for scattering off the elements, and this is mainly due to the fact that
the elements have multiple isotopes and completely destructive interference cannot be
simultaneously achieved for all isotopes~\cite{Feng11,Feng13a}. We would like to point
out that even for elements with only one naturally abundant isotope, one still cannot find
a $g_{\rm np}$ value such that $D_{\rm p} \rightarrow 0$ since the neutron form factor is
generally different from the proton form factor in a nucleus.

The above results indicate that, the form factor effect can not only significantly change
the sensitivities of the detectors, but also affect the interaction behaviors of the
target element with the isospin-violating DM in various isospin-violating regions we
are interested in for DM direct detection experiments. In future when different direct
detection experiments using various target elements had specify the DM signals
precisely, one may determine simultaneously the isospin-violation factor $g_{\rm np}$
and the form factor (and thus the neutron skin thickness as well as the symmetry energy),
especially if the IVDM particle has a larger mass of above tens of GeV. Conversely, our 
results indicate that precise knowledge on the symmetry energy or the neutron skin 
thickness (and thus the neutron and proton form factors) of $^{208}$Pb is of critical
importance for the direct detection of IVDM using Xe-based detectors. In particular,
the future experiment PREX-II~\cite{Pas12} at JLab is expected to improve significantly the
measurement accuracy of the neutron skin thickness for $^{208}$Pb and thus could make
important contribution to this issue.

\section{Conclusion}
\label{sec:conclusion}
In the present work, we have shown that isospin-violating dark matter (IVDM) indeed provides
a possible mechanism to ameliorate the tension among recent direct detection
experiments, including CDMS-II(Si), XENON100, LUX, and SuperCDMS(Ge).
For IVDM, we have demonstrated that the results of the DM direct detection
experiments based on neutron-rich target nuclei, e.g., Xe-based detector, may
strongly depend on the density slope $L(\rho_c)$ of the symmetry energy at a
subsaturation cross density $\rho_c \approx 0.11 $fm$^{-3}$, which is presently
largely unknown and uniquely determines the neutron skin thickness and thus the
relative difference of neutron and proton form factors of the target nuclei.

In particular, using the proton and neutron form factors obtained from
Skyrme-Hartree-Fock calculations by varying the $L(\rho_c)$ within the 
uncertainty region set by the latest model-independent measurement of the 
neutron skin thickness from PREX experiment at JLab, we have found that 
although the form factor effects on the extracted bounds on DM-proton cross 
sections are negligible in the direct detection for isospin-invariant DM, 
they could become critically important in the detection for IVDM. Especially, 
for IVDM with neutron-to-proton coupling ratio fixed to
$f_{\rm n}/f_{\rm p}=-0.7$ in the mass region constrained by CMDS-II(Si), the
form factor effect may enhance the sensitivity of Xe-based detectors (e.g.,
XENON100 and LUX) to the DM-proton cross section by a factor of $3$, compared
with the results using the empirical Helm nuclear form factors extracted from
charge distributions. This form factor effect can even enhance the
sensitivity by more than a factor of $10$ for such kind of IVDM with mass larger
than $80$ GeV.

Furthermore, we have found that the form factor effect can significantly
modify the recoil spectrum of Xe-based detectors for heavy IVDM with
$f_{\rm n}/f_{\rm p}=-0.7$. We have also studied how the form factor effects 
change with the variation of $f_{\rm n}/f_{\rm p}$ and found that the 
$f_{\rm n}/f_{\rm p}$ value maximumly suppressing the sensitivity of the
detector may depend on the form factor, and the form factor effects can 
either enhance or suppress the sensitivity of the detector relative to the 
Helm's case depending on the specific value adopted for $f_{\rm n}/f_{\rm p}$.
Our results imply that the precise determination of the symmetry energy or 
the neutron skin thickness (and thus the neutron and proton form factors) of 
$^{208}$Pb is extremely useful for the direct detection of IVDM based on 
detectors with neutron-rich targets (e.g., xenon).

\section*{Acknowledgments}

The authors would like to thank Fei Gao for useful discussions.
This work was supported in part by the NNSF of China under Grant Nos. 11275125
and 11135011, the Shanghai Rising-Star Program under grant No. 11QH1401100,
the ``Shu Guang'' project supported by Shanghai Municipal Education Commission
and Shanghai Education Development Foundation, the Program for Professor of
Special Appointment (Eastern Scholar) at Shanghai Institutions of Higher Learning,
and the Science and Technology Commission of Shanghai Municipality (11DZ2260700).


\appendix*

\section{An empirical parametrization for proton and neutron form factors}
\label{sec:appendix}
%

%
\begin{figure}[tbp]
\includegraphics[width=12 cm]{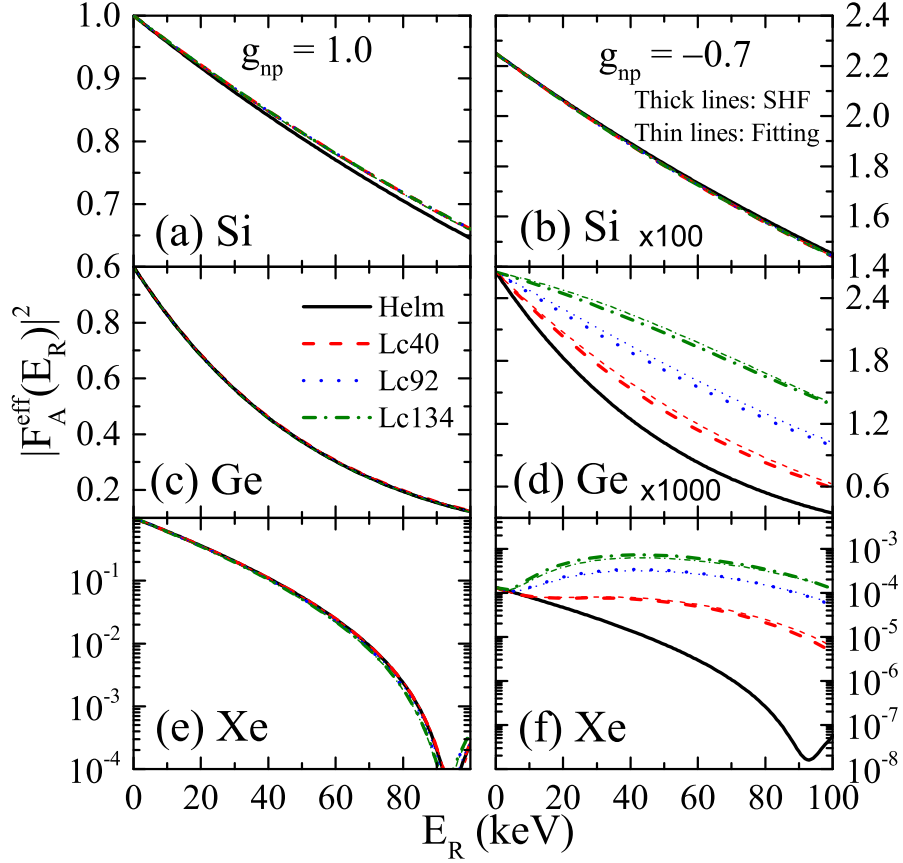}
\caption{(Color online) The averaged effective form factors as functions of recoil
energy $E_{\rm R}$ for Si (upper panels), Ge (middle panels) and Xe (lower panels) with
$g_{\rm np} = 1.0$ (left panels) and $g_{\rm np} = -0.7$ (right panels) obtained from
SHF with Lc40, Lc92 and Lc134 (thick lines) and from the empirical parametrization
for proton and neutron form factors in Eq.~(\ref{fitting}) (thin lines) with
$\Delta r_{\rm np} = 0.15 \, {\rm fm}$, $0.33 \, {\rm fm}$ and
$0.49 \, {\rm fm}$. The empirical Helm form
factor is also included for comparison.}
\label{ffs02}
\end{figure}

Shown in Fig.~\ref{fnp} are the proton and neutron form factors as functions
of momentum transfer for $^{28}$Si and $^{132}$Xe from SHF calculations with Lc40,
Lc92 and Lc134. The results for other isotopes of Si and Xe as well as Ge
isotopes can also be calculated in the same approach. Thanks to the
approximately linear relationship between the symmetry energy density slope
parameter $L(\rho_{\rm c})$ and the neutron skin thickness $\Delta r_{\rm np}$
of $^{208}$Pb, it is possible to approximate the proton and neutron form factors
obtained from SHF calculations with varied $L(\rho_{\rm c})$ by analytical
parameterizations with explicit dependence on the $\Delta r_{\rm np}$ of $^{208}$Pb.
In particular, here we construct the analytical expressions for the proton and neutron
form factors separately according to Helm-like form factor with an additional term
to consider isospin dependence and the symmetry energy effects (and thus
$\Delta r_{\rm np}$ of $^{208}$Pb), i.e.,
\begin{equation}
F_{\rm A}^{\rm n(p)}(q)
 =  3  \frac{j_{1}(q r_{\rm p(n)})}{q r_{\rm p(n)}} \times
e^{-(qs)^2/2} \, ,
\label{fitting}
\end{equation}
where the effective proton (neutron) distribution radius is parameterized as
\begin{equation}
r_{\rm p(n)}^2  =  c_{\rm p(n)}^2 + \frac{7}{3} \pi^2 a^2 - 5 s^2,
\label{rnp}
\end{equation}
with
\begin{eqnarray}
&c_{\rm p}&  =  r_{0}A^{\frac{1}{3}} + e - d \cdot\left(\Delta r_{\rm np}\right)^{\frac{1}{4}} \frac{Z(N-Z)}{A^2}, \label{cp}\\
&c_{\rm n}&  =  r_{0}A^{\frac{1}{3}} + e + d \cdot \Delta r_{\rm np} \frac{N(N-Z)}{A^2}.  \label{cn}
\end{eqnarray}
Compared with the parametrization of Helm form factor (i.e., Eq.~(\ref{helm})), the
parametrization here includes an additional term, namely, the last term in Eq.~(\ref{cp})
and Eq.~(\ref{cn}). In addition, our parametrization here distinguishes protons and
neutrons, and depends on explicitly the proton and neutron numbers of the nucleus and
the $\Delta r_{\rm np}$ of $^{208}$Pb.
%
\begin{table}
\begin{center}
\caption{Parameters in the empirical parametrization of
Eq. (\ref{fitting}) for proton and neutron form factors.}
\begin{tabular}{ccccc}
\hline\hline
 ~~~ &~~~~~& Proton &~~~~~& Neutron \\
\hline
 $r_0$ (fm)   &  & 1.2249  &  &  1.1425  \\
 $d$ (fm)           &  & 5.6794  &  &  5.4433  \\
 $e$ (fm)           &  & -0.0436 &  &  0.2737  \\
 $a$ (fm)           &  & 0.2898  &  &  0.2200  \\
 $s$ (fm)           &  & 0.8521  &  &  0.9752  \\
\hline\hline
\end{tabular}
\label{Table}
\end{center}
\end{table}

By fitting a number of proton and neutron form
factors calculated from SHF approach with varied symmetry energies, including the 9
stable Xe isotopes, 5 stable Ge isotopes and 3 stable Si isotopes with the symmetry
energy density slope parameter $L(\rho_{\rm c})$ varying from $40.5 \, {\rm MeV}$
to $134.2 \, {\rm MeV}$, we obtain all parameters in the above expressions and they
are listed in Table~\ref{Table}. In order to test the quality of the above parametrization,
we compare in Fig.~\ref{ffs02} the averaged effective form factors as functions of $E_{\rm R}$
for Si, Ge and Xe using the above parameterized proton and neutron form factors with
the results obtained from the SHF approach. It is very interesting to see that one can
indeed simply use just one set of parameters given in Table~\ref{Table} to well describe
the proton and neutron form factors in different nuclei in a large mass region with
very different values of the symmetry energy density slope parameter $L(\rho_{\rm c})$
(i.e., $\Delta r_{\rm np}$ of $^{208}$Pb).


\end{document}